\documentstyle[preprint,aps,epsf]{revtex}
\begin{document}
\tightenlines
\preprint{\vbox{\baselineskip=14pt
    \rightline{UH-511-941-99} \break
    \rightline{September 1999}}}
\title{Neutrino Anomalies without Oscillations
\thanks{Invited
talk presented at ``Beyond the Desert 1999'', Castle, Ringberg,
Tegernsee, Germany, June 6-12, 1999.}}

\author{\vspace{.25in}
Sandip Pakvasa}

\address{
Department of Physics \& Astronomy, University of Hawaii,
Honolulu, HI 96822 USA}

\maketitle
\begin{abstract}
I review explanations for the three neutrino anomalies
  (solar, atmospheric and LSND) which go beyond the ``conventional''
  neutrino oscillations induced by mass-mixing.  Several of these
  require non-zero neutrino masses as well.
 \end{abstract} 

  \section{Introduction}
  
As is well-known, it is not possible to account for all three neutrino
anomalies with just the three known neutrinos $(\nu_e, \nu_\mu$ and
$\nu_\tau)$. If one or more of them can be explained 
in some other way, then no extra sterile neutrinos
need be invoked.  This is one motivation for exotic scenarios. In any
case it is important to rule out all explanations other
than oscillations in order to establish neutrino mixing and
oscillations as the unique explanation for the three observed neutrino
anomalies.

I should mention that, in general, some (but not all) non-oscillatory
explanations of the neutrino anomalies will involve non-zero neutrino
masses and mixings.  Therefore, they tend to be neither elegant nor
economical.  But the main issue here is whether we can establish
oscillations unequivocally and uniquely as the cause for the observed
anomalies.

I first summarize some of the exotic scenarios and then 
consider each anomaly in turn.

\section{Mixing and Oscillations of massless $\nu's$}

There are three different ways that massless neutrinos may mix and even
oscillate.  These are as follows.

1. If flavor states are mixtures of massless as well massive states, then
when they are produced in reactions with Q-values smaller than the
massive state; the flavor states are massless but not
orthogonal\cite{lee}.  
For example, if
$\quad  \nu_e = \sum_{i=1}^{4}U_{ei}\nu_i$, 
\noindent
$\nu_\mu=\sum_{i=1}^{4} U_{\mu i}
\nu_i$ and $m_i = 0$ for $i= 1$ to 3 but $m_4$ = 50 GeV; then ``$\nu_e$''
produced in $\beta$-decay and ``$\nu_\mu$'' produced in $\pi$-decay
are massless but not orthogonal and
\begin{equation}
\left < \nu_e \mid \nu_\mu \right >  =  - U_{e4}^* \ U_{\mu 4}
\end{equation}

On the other hand $\nu_e$ and $\nu_\mu$ produced in $W$ decay will be
not massless and will be more nearly orthogonal.  Hence, the definition of
flavor eigenstate is energy and reaction dependent and not fundamental.
Current limits\cite{langacker} on orthogonality of $\nu_e, \nu_\mu$ and $\nu_\tau$ make
it impossible for this to play any role in the current neutrino
anomalies.

2.  When Flavor Changing Neutral Currents as well as Non-Universal
    Neutral Current Couplings of neutrinos exist, for example:
\begin{eqnarray}
& &\epsilon_q \frac{4 G_F}{\sqrt{2}} \left \{ 
\bar{\nu}_{el} \ \gamma_\mu \ \nu{_{\tau_{L}}} \bar{q}_L \ \gamma_\mu q_L \ \ +
h.c. \right \}    \nonumber \\
& + & \epsilon'_q \frac{4 G_F}{\sqrt{2}} \ \left \{ 
\bar{\nu}_{e  L} \ \gamma_\mu \ \nu{_{e_{L}}} - \bar{\nu}_{\tau L} 
\gamma_{\mu} \ \nu_{\tau L} \right \}  \left \{ \bar{q}_L \ \gamma_\mu \ q_L
\right \}
\end{eqnarray}
Such couplings arise in R-parity violating supersymmetric theories\cite{roulet}.
In general, in these theories, one expects neutrino masses at some
level here it is assumed that the FCNC provides the dominant effect.
In this case the propagation of $\nu_e$ and $\nu_\tau$ in matter is
described by the equation:
\begin{eqnarray}
i \frac{d}{dt}
\left (
\begin{array}{c}
\nu_e \\
\nu_\tau
\end{array}
\right )_L 
=
\left (
\begin{array}{cc}
\alpha + \beta & \gamma \\
\gamma         & -\beta
\end{array}
\right )
\left (
\begin{array}{c}
\nu_e \\
\nu_\tau
\end{array}
\right )_L
\end{eqnarray}
where $\alpha =G_F N_e, \beta = \epsilon_q G_F N_q$ and
$\gamma = \epsilon'_q G_F N_q$.  There is a resonance at
$\alpha + 2 \beta = 0$    or $\epsilon_q = -\frac{1}{2}
N_e/N_q$
and $\nu_e$ can convert to $\nu_\tau$ completely.  This matter effect
which effectively mixes flavors in absence of masses was first pointed
out by Wolfenstein in the same paper\cite{wolfenstein} where matter 
effects were first discussed.  The resonant conversion is just like in the conventional
MSW effect, except that there is no energy dependence and $\nu$ and $\bar{\nu}$
are affected the same way.  This possibility has been discussed in connection with both
solar and atmospheric anomalies.

3.  (a) Flavor violating Gravity wherein it is proposed that
    gravitational couplings of neutrinos are flavor
    non-diagonal\cite{gasperini} and equivalence principle is violated.  For
    example, $\nu_1$ and $\nu_2$ may couple to gravity with different
    strengths:

\begin{equation}
H_{gr} = f_1 GE \phi + f_2 GE \phi
\end{equation}
where $\phi$ is the gravitational potential.  Then if $\nu_e$ and
$\nu_\mu$ are mixtures of $\nu_1$ and $\nu_2$ with a mixing angle
$\theta$, oscillations will occur with a flavor survival probability
\begin{equation}
P= 1- sin^2 2 \theta sin^2 ( \frac{1}{2} \delta f \phi EL)
\end{equation}
when $\phi$ is constant over the distance L and $\delta f=f_1-f_2$ is
the small deviation from universality of gravitational coupling.
Equivalence principle is also violated.

(b) Another possibility is violation of Lorentz invariance\cite{coleman} wherein all
particles have their own maximum attainable velocities (MAV) which are all
different and in general also different from speed of light:  then if
$\nu_1$ and $\nu_2$ are MAV eigenstates with MAV's $v_1$ and
$v_2$ and $\nu_e$ and $\nu_\mu$ are mixtures of $\nu_1$ and $\nu_2$
with mixing angle $\theta$,
the survival probability of a given flavor is
\begin{equation}
P=1-sin^2 2 \theta \ sin^2 \left (\frac{1}{2} \delta v EL \right )
\end{equation}
where $\delta v = v_1-v_2$.

As far as neutrino oscillations are concerned, these two cases are
identical in their dependence on LE instead of L/E as in the
conventional oscillations\cite{glashow}.

\section{Neutrino Decay}

Neutrino decay\cite{discussion} implies a non-zero mass difference between two neutrino
states and thus, in general, mixing as well.  I will consider here only
non-radiative decays.  We assume a
component of $\nu_\alpha,$ i.e., $\nu_2$, to be the only unstable state,
with a rest-frame lifetime $\tau_0$, and we assume two flavor mixing,
for simplicity:
\begin{equation}
\nu_\mu = cos \theta \nu_2 \ + sin \theta \nu_1
\end{equation}
with $m_2 > m_1$.  From Eq. (2) with an unstable $\nu_2$, the $\nu_\alpha$
survival probability is
\begin{eqnarray}
P_{\alpha \alpha} &=& sin^4 \theta \ + cos^4 \theta {\rm exp} (-\alpha L/E)
            \\ \nonumber
&+& 2 sin^2 \theta cos^2 \theta {\rm exp} (-\alpha L/2E)
            cos (\delta m^2 L/2E),
\end{eqnarray}
where $\delta m^2 = m^2_2 - m_1^2$ and $\alpha = m_2/ \tau_0$.
Since we are attempting to explain neutrino data without oscillations
there are two appropriate limits of interest.  One is when the $\delta
m^2$ is so large that the cosine term averages to 0.  Then the survival
probability becomes
\begin{equation}
P_{\mu\mu} = sin^4 \theta \ + cos^4 \theta {\rm exp} (-\alpha L/E)
\end{equation}
Let this be called decay scenario A.  The other possibility is when
$\delta m^2$ is so small that the cosine term is 1, leading to a
survival probability of 
\begin{equation}
P_{\mu \mu} = (sin^2 \theta + cos^2 \theta {\rm exp} (-\alpha L/2E))^2
\end{equation}
corresponding to decay scenario B.  Decay models for both kinds of
scenarios can be constructed; although they require fine tuning and are
not particularly elegant.

\section{LSND}

In the LSND experiment, what is observed is the following\cite{athana}.  In the decay
at rest (DAR) which is $\mu^+ \rightarrow e^+ \nu_e \bar{\nu}_\mu$ 
which should give a pure
$\nu_e$ signal, there is a flux of $\bar{\nu_e's}$ at a level of about
$3.10^{-3}$ of the $\nu_e's$.  (There is a similar
signal for $\nu'_es$ accompanying $\bar{\nu}'_es$ in the decay in flight
of $\mu^- \rightarrow e^- \bar{\nu}_e \nu_\mu)$.  Now this could be
accounted for without oscillations\cite{some} provided that the conventional decay mode
$\mu^+ \rightarrow e^+ \nu_e \bar{\nu}_\mu (\mu^- \rightarrow
e^- \bar{\nu}_e \nu_\mu)$ is accompanied by the rare mode 
$\mu^+ \rightarrow e^+ \bar{\nu}_e {\rm X} \ ( \mu^- \rightarrow e^- \nu_e
\bar{\rm X})$ at a level of branching fraction of $3.10^{-3}$.  Assuming X
to be a single particle, what can X be?  It is straight forward to
rule out X as being (i) $\nu_\mu$ (too large a rate for
Muonium-Antimuonium transition rate), (ii) $\nu_e$ (too large a rate
for FCNC decays of
$Z$ such as $Z \rightarrow \mu\bar{e} + \bar{\mu} e)$ and (iii)
$\nu_\tau$ (too large a rate for FCNC decays of $\tau$ such as $\tau
\rightarrow \mu ee)$.

The remaining possibilities for X are $\bar{\nu}_\alpha$ or
$\nu_{\rm{sterile}}$.
No simple models exist which lead to such decays.  Rather Baroque models
can be constructed which involve a large number of new particles\cite{grossman}.

Experimental tests to distinguish this rare decay possibility from the
conventional oscillation explanation are easy to state.  In the rare decay
case, the rate is constant and shows no dependence on L or E.

	One can also ask whether the $\bar{\nu}_e$ events seen in LSND
could have been caused by new physics at the detector; for example a
small rate for the ``forbidden'' reaction $\bar{\nu}_\mu + p \rightarrow
n + e^+$.  This is ruled out, since a fractional rate of $3.10^{-3}$ for
this reaction, leads (via crossing) to a rate for the decay mode $\pi^+
\rightarrow e^+ \nu_\mu$ in excess of known bounds.

\section{Solar Neutrinos}

Oscillations of massless neutrinos via Flavor Changing Neutral Currents
(FCNC) and Non Universal Neutral Currents (NUNC) in matter have
been considered\cite{roulet,bahcall} as explanation for the solar
neutrino observations, most
recently by Babu, Grossman and Krastev\cite{krastev}.  Using the most recent data from
Homestake, SAGE, GALLEX and Super-Kamiokande, they find good fits with
$\epsilon_u \sim 10^{-2}$ and $\epsilon'_u \sim 0.43$ or $\epsilon_d
\sim$ 0.1 to 0.01 with $\epsilon'_d \sim 0.57$.

Since the matter effect in this case is energy independent, the
explanation for different suppression for $^8B$, $^7Be$ and pp neutrinos
as inferred is interesting.  The differing suppression arises from the
fact the production region for each of these neutrinos differ in
electron and nuclear densities.  This solution resembles the large angle
MSW solution with the difference that the day-night effect is
energy-independent.

The massless neutrino oscillations described in Eq.(6) also offer a
possible solution to solar neutrino observations.  There are three
solutions.  Two are characterized by $\sin^2 2 \theta \sim 2.10^{-3},
\delta v/2 \sim 6.10^{-19}$ and $\sin^2 2 \theta \sim 0.7, \delta v/2
\sim 10^{-21}$; these are analogs of the small angle MSW and large
angle MSW solutions\cite{mansour}.  These are now ruled
out\cite{pantaleone} (for both $\nu_e-\nu_\mu$
and $\nu_e-\nu_\tau)$ by the recent NUTEV data\cite{CCFR}.  The third one at $\sin^2
2 \theta \sim 1$ and $\frac{\delta v}{2} \sim 10^{-24}$ (the analog of
the vacuum solution) is the only one allowed\cite{gago}.  
This possibility can be tested in Long Baseline experiments.

The possibility of solar neutrinos decaying to explain the discrepancy
is a very old suggestion\cite{pakvasa}. 
The most recent analysis of the current solar neutrino data finds that
no good fit can be found:
$U_{ei} \approx 0.6$ and $\tau_\nu$ (E=10
MeV) $\sim 6$ to 27 sec. come closest\cite{acker}. 
The fits become acceptable only if the suppression
of the solar neutrinos is energy independent as proposed by several
authors\cite{harrison} (which is possible if
the Homestake data are excluded from the fit).  The above conclusions are
valid for both the decay scenarios A as well as B.

\section{Atmospheric Neutrinos}

The massless FCNC scenario has been recently considered for the
atmospheric  neutrinos by Gonzales-Garcia et al.\cite{gonzalez} with the matter effect 
supplied by the earth.
Good fits were found for the partially contained and multi-GeV events
with $\epsilon_q \sim 1, \epsilon'_q \sim 0.02$ as well $\epsilon_q \sim
0.08, \epsilon_q' \sim 0.07$.  The fit is poorer when higher energy
events corresponding to up-coming muons are included\cite{lipari,fogli}.  The expectations
for future LBL experiments are quite distinctive:  for MINOS, one expects
$P_{\mu\mu} \sim 0.1$ and $P_{\mu\tau} \sim 0.9.$ 
The story for massless neutrinos oscillating via violation of
Equivalence Principle or Lorentz Invariance is very similar.  Assuming
large $\nu_\mu - \nu_\tau$ mixing, and
$\delta v/2 \sim 2.10^{-22},$ a good fit to the contained
events can be obtained\cite{foot}; but as soon as multi-GeV and thrugoing muon
events are included, the fit is quite poor\cite{lipari,fogli}.

Turning to neutrino decay scenario A, it was found that it 
is possible to choose $\theta$ and $\alpha$ to provide a
good fit to the Super-Kamiokande L/E distributions of $\nu_\mu$ events and
$\nu_\mu/\nu_e$ event ratio\cite{fukuda}.  
The best-fit values of the two parameters are $cos^2 \theta
\sim 0.87$   and $\alpha \sim 1 GeV/D_E,$ where
$D_E=12800$ km is the diameter of the Earth.  
This best-fit $\alpha$ value corresponds
to a rest-frame $\nu_2$ lifetime of
\begin{equation}
\tau_0 = m_2/\alpha \sim 
\frac{m_2}{(1 eV)} \times 10^{-10} s.
\end{equation}
However, it was then shown that the fit to the higher energy events in
Super-K (especially the upcoming muons) is quite poor\cite{lipari,fogli}.

In all these three cases, the reason that the inclusion of high energy
upcoming muon events makes the fits poorer is very simple. The upcoming
muons come from much higher energy $\nu_\mu's$ and although there is
some suppression, it is less than what is observed for lower energy
events at the same L (zenith angle).  This is in accordance with
expectations from conventional oscillations.  The energy dependence in
the above three scenarios is different and fails to account for the
data.  In the FCNC case there is no energy dependence and so the high
energy $\nu_\mu's$ should have been equally depleted, in the FV Gravity
(or Lorentz invariance violation) at high energies the oscillations
should average out to give uniform 50\% suppression and in the decay A
scenario due to time dilation the decay is suppressed and there is
hardly any depletion of $\nu_\mu's$.

Turning to decay scenario B, consider the following possibility\cite{barger}.  The
three weak coupling states $\nu_\mu, \nu_\tau, \nu_s$ (where $\nu_s$ is a
sterile neutrino) may be related to the mass eigenstates $\nu_2 \nu_3
\nu_4$ by the approximate mixing matrix.
\begin{equation}
\left( \begin{array}{c} \nu_\mu\\ \nu_\tau\\ \nu_s \end{array} \right) =
\left( \begin{array}{ccc}  \cos\theta& \sin\theta& 0\\
                          -\sin\theta& \cos\theta& 0\\
                           0& 0& 1
\end{array} \right)
\left( \begin{array}{c} \nu_2\\ \nu_3\\ \nu_4 \end{array} \right)
\label{eq:mixing}
\end{equation}
and the decay is $\nu_2 \to \bar\nu_4 + J$. The electron neutrino,
which we identify with $\nu_1$, cannot mix very much with the other
three because of the more stringent bounds on its couplings\cite{barger1},
and thus our preferred solution for solar neutrinos would be
small angle matter oscillations.

Then the $\delta m_{23}^2$ in Eq.~(1) is not
related to the $\delta m_{24}^2$ in the decay, and can be very small,
say $ < 10^{-4}
\rm\, eV^2$ (to ensure that oscillations play no role in the atmospheric
neutrinos). In that case, the oscillating term is 1 and $P(\nu_\mu\to
\nu_\mu)$ becomes
\begin{equation}
P(\nu_\mu\to \nu_\mu) = (\sin^2 \theta + \cos^2 \theta e^{- \alpha L/2E})^2
\end{equation}
This is identical to Eq.~(13) in Ref.~\cite{discussion}. 

In order to compare the  predictions of this model with the standard
$\nu_\mu \leftrightarrow \nu_\tau$ oscillation model,
we have  calculated  with Monte Carlo methods   the  event  rates
for contained, semi-contained  and upward-going  (passing and stopping)
muons  in the Super-K  detector,  in the  absence of `new  physics', and
modifying the muon neutrino  flux   according
to the   decay or  oscillation  probabilities   discussed  above.
We have then  compared our  predictions  with the SuperK data
\cite{fukuda},  calculating  a $\chi^2$  to
quantify the agreement   (or  disagreement)     between  data and
calculations.
In    performing  our  fit  (see Ref.~\cite{lipari}  for details)
we  do not take into account any  systematic  uncertainty, but we allow
the  absolute flux normalization to vary as  a free
parameter  $\beta$.

The  `no  new physics  model'     gives  a  very  poor  fit
to the data  with $\chi^2 = 281$ for
34  d.o.f.  (35  bins  and one   free parameter, $\beta$).
For the  standard $\nu_\mu \leftrightarrow \nu_\tau$ oscillation scenario
the  best fit    has  $\chi^2 = 33.3$   (32  d.o.f.)
and the values of the relevant parameters are
$\Delta m^2 = 3.2\times10^{-3}$~eV$^2$,
$\sin^2 2 \theta = 1$ and $\beta = 1.15$.
This  result is in  good  agreement  with the
detailed  fit  performed  by the SuperK collaboration \cite{fukuda}
giving  us  confidence   that   our  simplified    treatment
of detector acceptances  and  systematic uncertainties is
reasonable.
The decay  model of Equations (3) and (4) above   gives an equally good fit
with a minimum $\chi^2 = 33.7$ (32 d.o.f.)
for the choice  of  parameters
\begin{equation}
\tau_\nu/m_\nu = 63\rm~km/GeV,
\ \cos^2 \theta = 0.30
\end{equation}
and normalization $\beta = 1.17$.

In Fig.~1  we  compare  the
best fits of the two  models  considered
(oscillations  and  decay)  with the
SuperK  data.
In the figure we show
(as  data points  with statistical error bars)
the ratios between the SuperK data and the Monte Carlo  predictions
calculated  in the  absence of oscillations or other
form of `new physics' beyond the standard model.
In the six  panels  we  show   separately  the  data
on $e$-like and  $\mu$-like events in the sub-GeV and multi-GeV
samples,   and  on   stopping and passing
upward-going muon events.
The   solid (dashed) histograms
correspond to  the  best fits for the decay  model
($\nu_\mu \leftrightarrow \nu_\tau$ oscillations).
One  can  see  that the  best fits  of the two  models
are  of comparable  quality.
The reason  for the similarity  of the results  obtained
in  the two  models  can be understood by looking  at
Fig.~2, where  we show
the survival probability $P(\nu_\mu \to \nu_\mu)$
of muon neutrinos   as  a  function
of $L/E_\nu$ for  the  two  models   using the
best  fit  parameters.
In the case  of the neutrino  decay model   (thick  curve)
the probability   $P(\nu_\mu \to \nu_\mu)$
monotonically  decreases   from    unity  to  an  asymptotic  value
$\sin^4 \theta \simeq  0.49$.
In the case of  oscillations the  probability  has  a sinusoidal
behaviour  in $L/E_\nu$.  The  two  functional    forms
seem     very different;  however,  taking  into  account  the
resolution in $L/E_\nu$,  the  two  forms
are  hardly  distinguishable.
In fact, in the    large $L/E_\nu$    region, the oscillations
are  averaged  out  and the survival  probability there
can  be  well  approximated  with 0.5  (for  maximal  mixing).
In  the region  of  small  $L/E_\nu$  both probabilities  approach
unity.
In the region $L/E_\nu$ around  400~km/GeV, where  the  probability for the
neutrino oscillation model  has the first  minimum,
the  two  curves are  most  easily  distinguishable, at least in
principle.

\medskip\noindent{Decay Model}

There are two decay possibilities that can be considered: (a)~$\nu_2$ decays to
$\bar\nu_4$  which is dominantly $\nu_s$ with $\nu_2$ and $\nu_3$ mixtures of
$\nu_\mu$ and $\nu_\tau$, as in Eq.~(\ref{eq:mixing}), and
(b)~$\nu_2$ decays into $\bar\nu_4$ which is dominantly $\bar\nu_\tau$ and
$\nu_2$
and
$\nu_3$ are mixtures of $\nu_\mu$ and $\nu_s$.
In both cases the decay interaction has to be of the form
\begin{equation}
{\cal{L}}_{int} = g_{24} \ \overline{\nu_{4_{L}}^c} \ \nu_{2_{L}} J + h.c.
\end{equation}
where $J$ is a Majoron field that is dominantly iso-singlet (this avoids any
conflict with the invisible width of the $Z$).  Viable
models for both the above cases can be constructed \cite{valle,joshipura}.
However, case (b) needs additional iso-triplet light scalars which cause
potential problems with Big Bang Nucleosynthesis (BBN), and there is some
preliminary evidence from SuperK against $\nu_\mu$--$\nu_s$ mixing
\cite{kajita}. Hence we
only consider case (a), i.e.\ $\nu_2\to\bar\nu_4 + J$ with $\nu_4\approx
\nu_s$, as implicit in Eq.~(\ref{eq:mixing}).
With this interaction, the $\nu_2$ rest-lifetime is given by
\begin{equation}
\tau_2 = \frac
{16 \pi}{g^2} \cdot \
\frac{m_2}{\delta m^2 (1 + x)^2},
\end{equation}
where $\delta m^2 = m^2_2 - m^2_4$ and $x=m_4/m_2 \   (0 < x <1)$.
{}From the value of $\alpha^{-1} = \tau_2/m_2 = 63$~km/GeV found in the fit
and for $x=0$, we
have
\begin{equation}
g^2 \delta m^2 \simeq 0.16\rm\, eV^2
\end{equation}
Combining this with the bound on $g^2$ from $K \rightarrow \mu$ decays of $g^2
< 2.4\times10^{-4}$ \cite{barger1} we have
\begin{equation}
\delta m^2 > 650 \rm\ eV^2 \,.
\end{equation}
Even with a generous interpretation of the uncertainties in the fit,
this $\delta m^2$ implies a minimum mass difference in the range of about
25~eV.
Then $\nu_2$ and $\nu_3$ are nearly degenerate with masses
$\stackrel{\sim}{>} {\cal O}$(25~eV) and $\nu_4$ is relatively light. We assume
that
a similar coupling of $\nu_3$ to $\nu_4$ and J is somewhat weaker
leading to a significantly longer lifetime for $\nu_3$, and the instability of
$\nu_3$ is irrelevant for the analysis of the atmospheric neutrino
data.

For the atmospheric neutrinos in SuperK, two kinds of tests have been proposed
to distinguish between $\nu_\mu$--$\nu_\tau$ oscillations and
$\nu_\mu$--$\nu_s$ oscillations. One is based on the fact that matter effects
are present for $\nu_\mu$--$\nu_s$ oscillations\cite{bdppw} but are nearly absent for
$\nu_\mu$--$\nu_\tau$ oscillations\cite{panta} leading to differences in the zenith angle 
distributions  due to
matter effects on upgoing neutrinos \cite{lipari2}.
The other is the fact that the neutral current rate
will be affected in $\nu_\mu$--$\nu_s$ oscillations but not for
$\nu_\mu$--$\nu_\tau$ oscillations as can be measured in  events
with single $\pi^0$'s \cite{smirnov}. In these tests our decay scenario will
behave
as a hybrid in that there is no matter effect but there is some effect in
neutral current rates.

\medskip
\noindent{Long-Baseline Experiments}

The survival probability of $\nu_\mu$ as a function of $L/E$ is given in
Eq.~(1). The conversion probability into $\nu_\tau$ is given by
\begin{equation}
P(\nu_\mu\to\nu_\tau) = \sin^2\theta \cos^2\theta (1-e^{-\alpha L/2E})^2 \,.
\end{equation}
This result differs from $1-P(\nu_\mu\to\nu_\mu)$ and hence is different from
$\nu_\mu$--$\nu_\tau$ oscillations. Furthermore, $P(\nu_\mu\to\nu_\mu)
+ P (\nu_\mu\to \nu_\tau)$ is
not 1 but is given by
\begin{equation}
P (\nu_\mu\to\nu_\mu) + P(\nu_\mu\to\nu_\tau) = 1 - \cos^2\theta (1 -
e^{-\alpha L/E})
\end{equation}
and determines the amount by which the predicted neutral-current rates are
affected compared to the no oscillations (or the $\nu_\mu$--$\nu_\tau$
oscillations) case.
In Fig.~3 we give the results for $P(\nu_\mu\to\nu_\mu)$,
$P(\nu_\mu\to\nu_\tau)$ and $P(\nu_\mu\to\nu_\mu) +
P(\nu_\mu\to\nu_\tau)$ for the decay model and compare them to the
$\nu_\mu$--$\nu_\tau$ oscillations, for both the K2K\cite{who} and
MINOS\cite{minos} (or the corresponding European project\cite{NGS})
long-baseline experiments, with the oscillation and decay parameters as
determined in the fits above.

The K2K experiment, already underway, has a low energy beam $E_\nu
\approx 1\mbox{--}2$~GeV and a baseline $L=250$~km.  The MINOS experiment will have
3 different beams, with average energies $E_\nu = 3,$ 6 and 12 GeV and a
baseline $L=732$~km.  The approximate $L/E_\nu$ ranges are thus 125--250~km/GeV for
K2K and 50--250~km/GeV for MINOS.  The comparisons in Figure 3 show that the
energy dependence of $\nu_\mu$ survival probability and the neutral
current rate can both distinguish between the decay and the oscillation
models.  MINOS and the European project may also have $\tau$ detection
capabilities that would allow additional tests.

\medskip
\noindent{Big Bang Nucleosynthesis}

The decay of $\nu_2$ is sufficiently fast that all the neutrinos ($\nu_e,
\nu_\mu, \nu_\tau, \nu_s$) and the Majoron may be expected to equilibrate in
the early universe before the primordial neutrinos decouple. When they achieve
thermal equilibrium each Majorana neutrino contributes $N_\nu = 1$ and the
Majoron contributes $N_\nu = 4/7$ \cite{kolb}, giving and effective number of
light
neutrinos $N_\nu = 4{4\over7}$ at the time of Big Bang Nucleosynthesis. From
the observed primordial abundances of $^4$He and $^6$Li, upper limits on
$N_\nu$ are inferred, but these depend on which data are
used\cite{olive,lisi,burles}. Conservatively, the upper limit to $N_\nu$ could
extend up to 5.3 (or even to 6 if $^7$Li is depleted in halo stars\cite{olive}).

\medskip
\noindent{Cosmic Neutrino Fluxes}

Since we expect both $\nu_2$ and $\nu_3$ to decay, neutrino beams
from distant sources (such as Supernovae, active galactic nuclei and
gamma-ray bursters) should contain only $\nu_e$ and $\bar\nu_e$
but no $\nu_\mu$, $\bar\nu_\mu$, $\nu_\tau$ and $\bar\nu_\tau$.
This is a very strong prediction of our decay scenario.
We can compare the very different expectations for neutrino flavor mixes
from very distant sources such as AGN's or GRB's. Let us suppose that at the
source the flux ratios are typical of a beam dump, a reasonable assumption:
$N_{\nu_{e}}:N_{\nu_{\mu}}:N_{\nu_{\tau}} = 1:2:0$.  Then, for the
conventional oscillation scenario, when all the $\delta m^2$'s satisfy
$\delta m^2 \ L/4E >>1$, it turns out curiously enough that for
a wide 
variety of choices of neutrino mixing matrices, the final flavor mix is the
same, namely: $N_{\nu_{e}}:
N_{\nu_{\mu}}:N_{\nu_{\tau}} = 1:1:1$.  In the  case of the decay B
scenario, as mentioned here, we have
$N_{\nu_{e}}:N_{\nu_{\mu}}:N_{\nu_{\tau}} = 1:0:0.$ The two are quite distinct.  Techniques for
determining these flavor mixes in future KM3 neutrino telescopes have been
proposed\cite{learned}.

\medskip
\noindent{Reactor and Accelerator Limits}

The $\nu_e$ is essentially decoupled  from the decay state $\nu_2$ so the null
observations from the CHOOZ reactor are satisfied\cite{chooz}. The mixings of $\nu_\mu$ and
$\nu_\tau$ with $\nu_s$ and $\nu_e$ are very small, so there is no conflict
with stringent accelerator limits on flavor oscillations with large $\delta
m^2$~\cite{zuber}.

\medskip

 In summary, neutrino decay remains a viable
 alternative to neutrino oscillations as an explanation of the atmospheric
 neutrino anomaly. The model consists of two nearly degenerate mass
 eigenstates $\nu_2$, $\nu_3$ with mass separation $\stackrel{\sim}{>} {\cal O}
(25$~eV) from
 another nearly degenerate pair $\nu_1$, $\nu_4$.
The $\nu_\mu$ and $\nu_\tau$ flavors are approximately composed of
$\nu_2$ and $\nu_3$, with a mixing angle $\theta_{23} \simeq 57^\circ$.
The state $\nu_2$ is unstable, decaying to $\bar{\nu}_{4}$ and a Majoron
with a lifetime $\tau_2 \sim 10^{-12}$ sec.  The electron neutrino
$\nu_e$ and a sterile neutrino $\nu_s$ have negligible mixing with $\nu_\mu,
\nu_\tau$ and are approximate mass eigenstates ($\nu_e \approx \nu_1,
\nu_s \approx \nu_4)$, with a small mixing angle $\theta_{14}$ and a
$\delta m_{41}^2 \approx10^{-5}\rm\,eV^2$ to explain the solar
neutrino anomaly.
The states $\nu_3$ and $\nu_4$ are also unstable, but with $\nu_3$ lifetime
somewhat longer and $\nu_4$ lifetime much longer than the
$\nu_2$ lifetime.
This decay
scenario is difficult to distinguish from oscillations because of the
smearing in both L and $E_\nu$ in atmospheric neutrino events.  However,
long-baseline experiments, where $L$ is fixed, should be able to establish
whether the dependence of $L/E_\nu$ is exponential or sinusoidal. In
our scenario only $\nu_1$ is stable.  Thus, neutrinos of supernovae
or of extra galactic origin would be almost entirely $\nu_e$.
The contribution of the electron neutrinos and the Majorons to the cosmological
energy density $\Omega$ is negligible and  not relevant for large
scale structure formation.

Another proposal for explaining the atmospheric neutrinos is based on
decoherence of the $\nu_\mu's$ in the flux\cite{grossman1}.  The idea is that
$\nu_\mu's$ are interacting and getting tagged before
their arrival at the detector.  The cause is unknown but could be a
number of speculative possibilities such as a large neutrino background,
new flavor sensitive interactions in an extra dimension, violation of
quantum mechanics etc.  The $\nu_\mu$ survival probability goes as:
\begin{equation}
P= \frac{1}{2} \left [
1+ cos 2 \theta exp (-t/\tau) \right ]
\end{equation}
The Super-Kamiokande data can be fit by choosing $\tau \sim 10^{-2} s$
and $sin 2 \theta \sim 0.4$. A detailed fit to all the data over the
whole energy range has not been attempted yet.

\section{Conclusion}

As I mentioned at the beginning, the main motivation for this exercise
is to try to establish neutrino oscillations (due to mass-mixing) as the
unique explanation of the observed anomalies.  Even if neutrinos have
masses and do mix, the observed neutrino anomalies may not be due to
oscillations but due to other exotic new physics.  These possibilities
are testable and should be ruled out by experiments.  I have tried to show
that we are beginning to carry this program out.

\section{Acknowledgments}

I thank Professor Klapdor-Kleingrothaus for the invitation to this
wonderful castle and for the hospitality here.
I thank Andy Acker, Vernon Barger, Yuval Grossman, Anjan
Joshipura, Plamen Krastev, John Learned, Paolo Lipari, Eligio Lisi,
Maurizio Lusignoli and
Tom Weiler for many enjoyable discussions and collaboration. This
work is supported in part by U.S.D.O.E. under grant DE-FG-03-94ER40833.


\begin{figure}
\vspace{.5in}
\centering\leavevmode
{\epsfxsize=3.5truein\epsffile{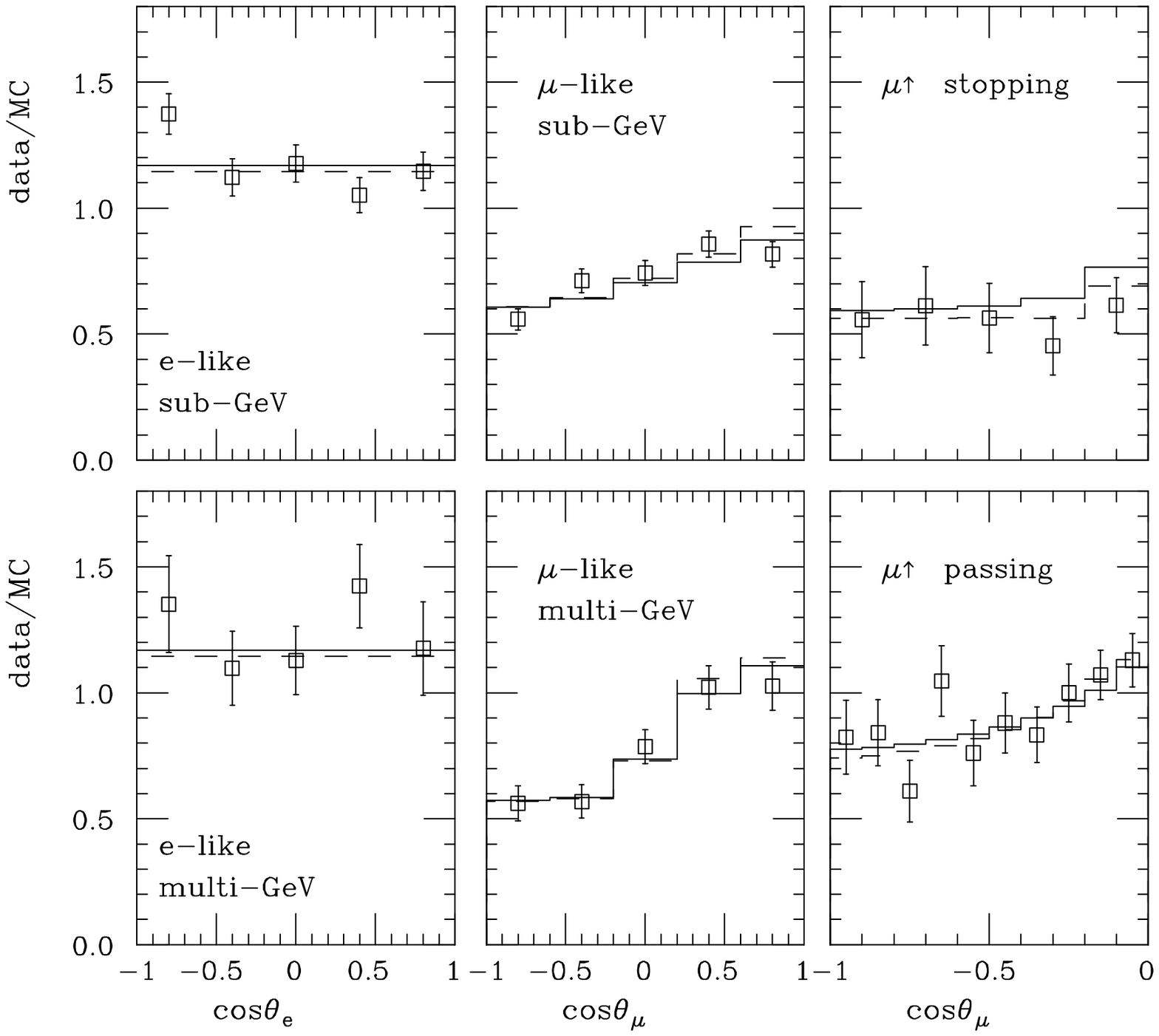}}
\caption[]{Comparison of decay model (solid histograms) and
$\nu_\mu$--$\nu_\tau$ oscillation model (dashed histograms) with SuperK data.}
\end{figure}

\begin{figure}
\centering\leavevmode
\epsfxsize=6.25in \epsffile{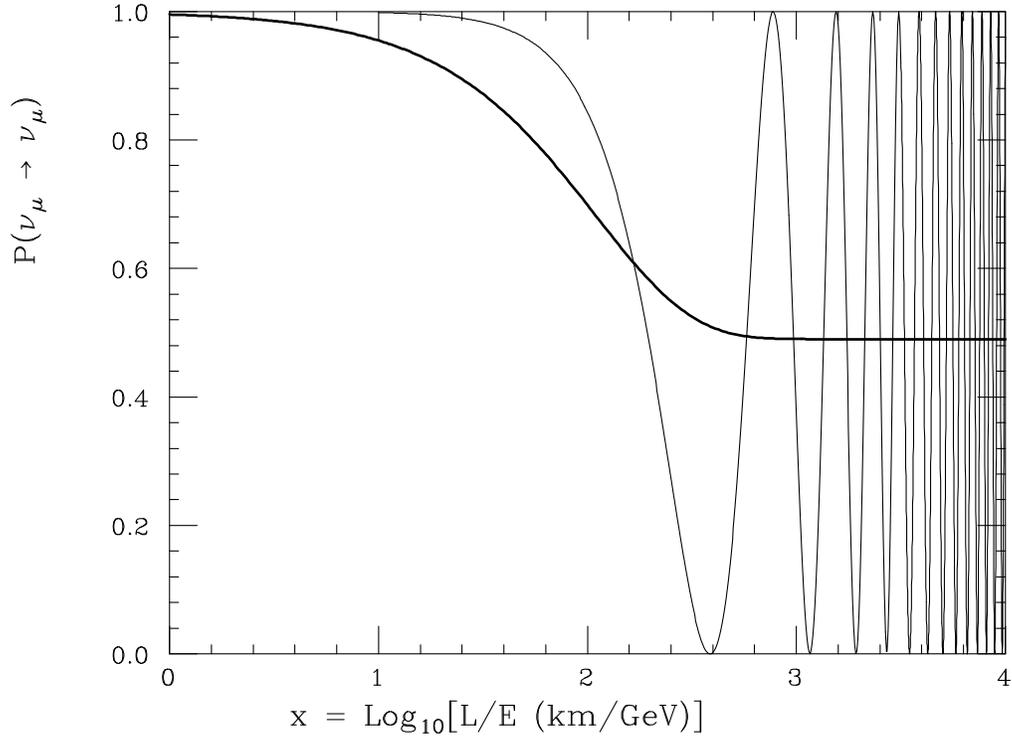}

\vspace{.5in}

\caption[]{Survival probabiliity for $\nu_\mu$ versus $\log_{10}(L/E)$ for the
decay model (heavy solid curve) and $\nu_\mu$ oscillation model (thin curve).}
\end{figure}

\begin{figure}
\centering\leavevmode
\epsfxsize=5.5in\epsffile{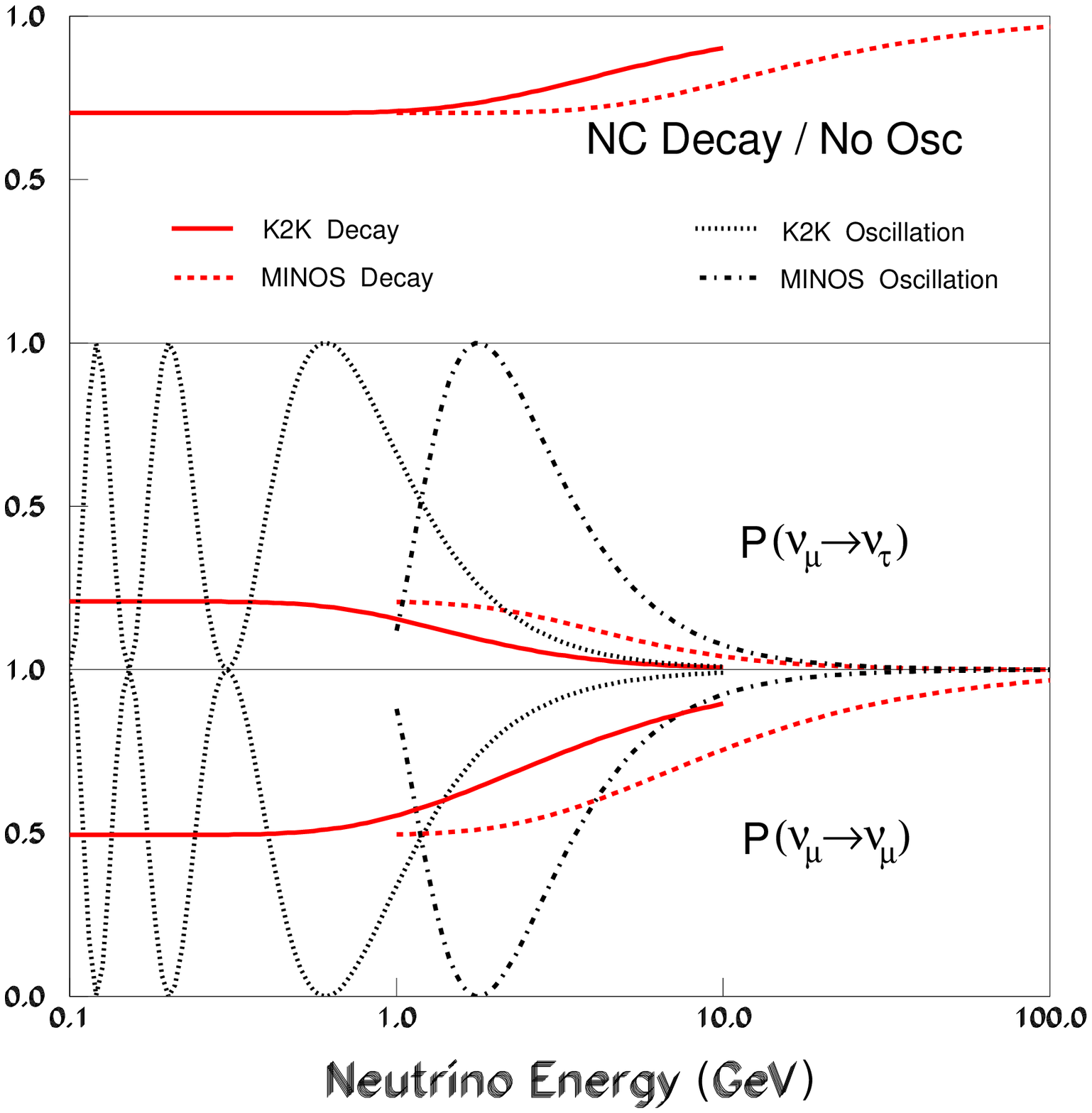}
\vspace{.5in}

\caption[]{Long-baseline expectations for the K2K and MINOS long-baseline
experiments from the
decay model and the $\nu_\mu$--$\nu_\tau$ oscillation model. The upper panel
gives the
neutral current predictions compared to no oscillations (or
$\nu_\mu$--$\nu_\tau$ oscillations).}
\end{figure}


\begin{thebibliography}{99}
 \bibitem{lee} B. W. Lee, S. Pakvasa, R. Shrock and H. Sugawara,
{\it Phys. Rev. Lett.} {\bf 38} (1977) 937;
S. Treiman. F. Wilczek and A. Zee, {\it Phys. Rev.} {\bf D16} (1977)
152.
  
\bibitem{langacker} P. Langacker and D. London, {\it Phys. Rev.}
{\bf D38} (1988) 907, S. Bergmann and A. Kagan, {\it Nucl. Phys.}
{\bf B358} (1999) 368.

\bibitem{roulet}E. Roulet, {\it Phys. Rev.} {\bf D44} (1991) 935;
M. M. Guzzo, A. Masiero and S. Petcov, {\it Phys. Lett.} {\bf B260} (1991)
154; V. Barger, R. J. N. Philips and K. Whisnant, {\it Phys. Rev.} {\bf D44} (1991) 1629.

\bibitem{wolfenstein}L. Wolfenstein, {\it Phys. Rev.} {\bf D17}
(1978) 2369.  

\bibitem{gasperini}M. Gasperini, {\it Phys. Rev.} {\bf D38} (1988) 2635;
A. Halprin and C. N. Leung, {\it Phys. Rev. Lett.} {\bf 67} (1991)
1833.

\bibitem{coleman}S. Coleman and S. L. Glashow, {\it Phys. Lett.} 
{\bf B405} (1997) 249.

\bibitem{glashow}S. Glashow, A. Halprin, P. I. Krastev, C. N. Leung and
J. Pantaleone, {\it Phys. Rev.} {\bf D56} (1977)
2433.

\bibitem{discussion}This discussion follows V. Barger, J. G. Learned,
S. Pakvasa and T. J. Weiler, {\it Phys. Rev. Lett.} {\bf 82} (1999)
2640.

\bibitem{athana} C. Athanassopoulos et al., the LSND Collaboration,
{\it Phys. Rev. Lett.} {\bf 77} (1996) 3082; 
{\it ibid} {\bf 81} (1998)
1774.

\bibitem{some}Some discussion of these scenarios can be found in
S. Bergmann and Y. Grossman, {\it Phys. Rev.} {\bf D59} (1999) 093005;
L. M. Johnson and D. McKay, {\it Phys. Lett.} {\bf B433} (1998) 355 and
P. Herczeg, {\it Proceedings of International Conference on Particle
Physics Beyond The Standard Model}, Jun 8-14, 1997; Castle Ringberg,
Germany; ed. by H. V. Klapdor-Kleingrothaus, (1998) 124.

\bibitem{grossman}Y. Grossman, (private communication).

\bibitem{bahcall}J. N. Bahcall and P. Krastev, hep-ph/9703267;
S. Bergmann, {\it Nucl. Phys.} {\bf B515} (1998) 363.

\bibitem{krastev}P. Krastev, (private communication).

\bibitem{mansour}S. W. Mansour and T-K. Kuo, hep-ph/9810510; see also
A. Halprin, C. N. Leung and J. Pantaleone, {\it Phys. Rev.} {\bf D53}
(1996) 5365; J. N. Bahcall, P. Krastev and C. N. Leung, {\it Phys. Rev.} 
{\bf D52} (1996) 1770; H. Minakata and H. Nunokawa, {\it Phys. Rev.} {\bf
D51} (1995) 6625.

\bibitem{pantaleone} J. Pantaleone, T. K. Kuo and S.W. Mansour,
hep-ph/9907478.


\bibitem{CCFR} The CCFR Collaboration, A. Romosan, et al, {\it
Phys. Rev.} {\bf D59} (1999) 031101, {\it Phys. Rev. Lett.} {\bf 78}
(1997) 2912.

\bibitem{gago} A.M. Gago, H. Nunokawa and R. Zukanovich-Funchal, hep-ph/9909250. 

\bibitem{pakvasa}S. Pakvasa and K. Tennakone, {\it Phys. Rev. Lett.}
{\bf 28} (1972) 1415; J. N. Bahcall, N. Cabibbo and A. Yahil, {\it
Phys. Rev. Lett.} {\bf 28} (1972) 316; See also A. Acker, A. Joshipura
and S. Pakvasa, {\it Phys. Lett.} {\bf B285} (1992) 371; Z. Berezhiani,
G. Fiorentini, A. Rossi and M. Moretti, {\it JETP Lett.} {\bf 55} (1992)
151.

\bibitem{acker}A. Acker and S. Pakvasa (in preparation); A. Acker and
S. Pakvasa, {\it Phys. Lett.} {\bf B320} (1994) 320.

\bibitem{harrison}  P. F. Harrison, D. H. Perkins and W. G. Scott,
hep-ph/9904297;
A. Acker, J. G. Learned, S. Pakvasa and T. J. Weiler, {\it Phys. Lett.} 
{\bf B298} (1993) 149; G. Conforto, M. Barone and C. Grimani, 
{\it Phys. Lett.} {\bf B447} (1999) 122; A. Strumia, {\it JHEP}
 {\bf 9904} (1999) 026.

\bibitem{gonzalez}M. C. Gonzales-Garcia, M. M. Guzzo, P. I. Krastev,
H. Nunokawa, O. Peres, V. Pleitez, J. Valle and R. Zukanovich Funchal, 
{\it Phys. Rev. Lett.} {\bf 82} (1999) 3202; See also F. Brooijmans, hep-ph/9808498.

\bibitem{lipari} P. Lipari and M. Lusignoli, {\it Phys. Rev.} {\bf D60}
(1999) 013003.

\bibitem{fogli} G. L. Fogli, E. Lisi and A. Marrone, hep-ph/9904248; G. L. Fogli,
E. Lisi and A. Marrone, {\it Phys. Rev.} {\bf D59} (1999) 117303;
S. Choubey and S. Goswami, hep/ph-9904257.

\bibitem{foot}R. Foot, C. N. Leung and O. Yasuda, {\it Phys. Lett.} {\bf
B443} (1998) 185.  

\bibitem{fukuda} Y.~Fukuda et al.(the Super-Kamiokande Collaboration), {\it 
Phys. Rev. Lett.} {\bf81} (1998) 1562; {\it ibid}, {\bf 82} (1998) 2644;
T. Kajita hep-ex/9810001. The use of more recent data(K. Scholberg,
hep-ex/9905016) does not change the conclusions.

\bibitem{barger} V. Barger, J. G. Learned, P. Lipari, M. Lusignoli,
S. Pakvasa and T. J. Weiler, hep-ph/9907421(Phys. Lett. B in press).

\bibitem{barger1} V. Barger, W-Y. Keung and S. Pakvasa, {\it Phys. Rev.}
{\bf D25} (1982) 907.

\bibitem{valle}J. Valle, {\it Phys. Lett.} {\bf 131B} (1983) 87;
G. Gelmini and J. Valle, {\it ibid} {\bf B142} (1983) 181; K. Choi and
A. Santamaria, {\it Phys. Lett. }{\bf B267} (1991) 504; A. Joshipura and
S. Rindani, {\it Phys. Rev.} {\bf D46} (1992) 300.

\bibitem{joshipura} A. Joshipura (private communication).

\bibitem{kajita} T. Kajita, Super-Kamiokande results presented at the
``Beyond the Desert'' Workshop, Castle Ringberg, Tegernsee, Germany,
June 6--12, 1999 (to be published in the proceedings).

\bibitem{bdppw} V. Barger, N. Deshpande, P.~Pal, R.J.N.~Phillips and
K.~Whisnant, {\it Phys. Rev.} {\bf D43}, 1759 (1991); E. Akhmedov, P. Lipari and M.
       Lusignoli, {\it Phys. Lett.} {\bf B300}, 128 (1993).

\bibitem{panta} J. Pantaleone, {\it Phys. Rev.} {\bf D49}, 2152 (1994).

\bibitem{lipari2}Q.~Liu and A.~Smirnov, {\it Nucl. Phys.} {\bf B524}, 505 (1998);
 P. Lipari and M. Lusignoli, {\it Phys. Rev.} {\bf D58}, 073005 (1998).

\bibitem{smirnov} F.~Vissani and A.~Smirnov, {\it Phys. Lett.} {\bf B432}, 376 (1998);
J.~Learned, S.~Pakvasa, and J.~Stone, {\it Phys. Lett.} {\bf B435}, 131 (1998);
L.~Hall and H.~Murayama, {\it  Phys. Lett.} {\bf B436}, 323 (1998).

\bibitem{who} KEK-PS E362, INS-924 report (1992).

\bibitem{minos} MINOS Collaboration, NuMI-L-375 report (1998).

\bibitem{NGS} NGS report, CERN 98-02, INFN/AE-98/05 (1998).

\bibitem{kolb} R. Kolb and M. S. Turner, {\bf The Early Universe}, Addison -
Wesley (1990).

\bibitem{olive} K. Olive and D. Thomas, hep-ph/9811444.

\bibitem{lisi} E. Lisi, S. Sarkar, F. Villante, {\it Phys. Rev.} {\bf D59},
123520 (1999).

\bibitem{burles} S. Burles, K. Nollett, J.  Truran and M. S. Turner,
Phys. Rev. Lett. {\bf 82}, 4176 (1999).

\bibitem{learned} J. G. Learned and S. Pakvasa, {\it Astropart. Phys.}
{\bf 3}, (1995) 267; F. Halzen and D. Saltzberg, {\it Phys. Rev. Lett.}
{\bf 81} (1998) 5722.

\bibitem{chooz}M.~Apollonio et al. (the CHOOZ Collaboration),  Phys. Lett. {\bf B420}, 320
(1998).

\bibitem{zuber} For a review of accelerator limits, see K.~Zuber, {\it Phys. 
Rep.} {\bf 305}, 295 (1998).

\bibitem{grossman1}Y. Grossman and M. Worah, hep-ph/9807511.
 \end{thebibliography}
\end{document}